\documentclass[12pt,preprint]{aastex}

\newcommand{\eb}{\begin{equation}}
\newcommand{\ee}{\end{equation}}

\shorttitle{Distortions of Astrometric Instruments}
\shortauthors{Makarov et al.}
\begin{document}

\title{The Worst Distortions of Astrometric Instruments and Orthonormal Models for Rectangular
Fields of View} 
\author{Valeri V. Makarov, Daniel R. Veillette, Gregory S. Hennessy, \& Benjamin F. Lane}
\affil{United States Naval Observatory, 3450 Massachusetts Ave. NW, Washington DC, 20392-5420}
\email{vvm@usno.navy.mil}

\begin{abstract}
The non-orthogonality of algebraic polynomials of field coordinates traditionally used to
model field-dependent corrections to astrometric measurements, gives rise to subtle adverse
effects. In particular, certain field dependent perturbations in the observational
data propagate into the adjusted coefficients with considerable magnification.
We explain how the worst perturbation, resulting in the largest solution error,
can be computed for a given non-orthogonal distortion model. An algebraic distortion model
of full rank can be converted into a fully orthonormal model based on the
Zernike polynomials for a circular field of view, or a basis of functions constructed
from the original model by a variant of the Gram-Schmidt orthogonalization process
for a rectangular field of view. The relative significance of orthonormal distortion
terms is assessed simply by the numerical values of the corresponding coefficients.
Orthonormal distortion models are easily extendable when the distribution of
residuals indicate the presence of higher order terms.
\end{abstract}

\keywords{astrometry --- methods: data analysis}
\section{Introduction}
\label{firstpage}
\label{Introduction}
The need to exactly characterize the mapping of celestial angular coordinates onto  Cartesian coordinates
in the plane of a detector, expressed in suitable units, arises in many astronomical observations.
Often called distortion models, such mapping models usually include the basic geometry of an
optical projection, possible deviations in the focal length, optical aberrations, etc. The propagation
of accidental error in traditional distortion models based on algebraic polynomials of field coordinates
has been studied in depth, and the use of orthogonal functions has been advocated in the astrometric literature.
However, it appears that simple algebraic polynomials are still in use for large-volume and
high-precision data reduction systems. For example, \citet{mca} represent the field-dependent errors in the
FGS observations with the HST with two independent fifth order polynomials for $x$ and $y$, a total
of 16 terms in each polynomial. The distortion model is expressed through 32 coefficients, which were
initially determined by ray-traces on the ground, but later updated several times by means of special
calibration observations of M35 while in-orbit. The high-volume astrometric data reductions in the 
Sloan Digital 
Sky Survey \citep[SDSS,][]{pier} were performed with a two-step characterization of field distortions.
For each CCD separately, the geometric part of the coordinate transformation was modeled as a
third order polynomial of $y$ only, because the observations were taken in the time-delayed integration mode.
This transformation was further superimposed with a 2D affine transformation. \citet{bel} used independent 3rd order algebraic polynomials for $\Delta x$ and $\Delta y$ to fit field-dependent corrections to a
nominal distortion model (sometimes called astrometric transfer function), represented by a conformal
transformation of field coordinates.

We investigate the propagation of systematic error and conclude that the use of non-orthogonal
terms can lead to significant errors in the instrument characterization. The worst systematic perturbation
in the data, producing the largest possible error in the model parameters, can be exactly computed
for any particular model, which is exemplified by the JMAPS astrometric project currently under development
\citep{hen}.
For the original JMAPS focal plane model, the  worst perturbation is amplified by a factor of 7 with respect
to the more benign perturbations of the same norm. To avoid this potentially harmful build-up of systematic error,
we suggest using orthogonal functions on the unit square (for a square detector) and explain how an existing
non-orthogonal model can be orthogonalized through the Gram-Schmidt process. Determining the significance
and temporal stability  of individual model terms, which are nearly independent on sufficiently
dense star fields, becomes straightforward.

\section{Polynomial plate models}
Traditionally, the  transformation of standard coordinates $\xi$, $\eta$  (related to the local angular coordinates on the
celestial sphere through a simple gnomonic projection) into Cartesian coordinates in the detector plane
$x$, $y$ is represented by a simple algebraic polynomial in powers of $x$ and $y$. The repercussions
of this choice for the propagation of accidental measurement errors into the astrometric plate reductions
have been discussed in numerous papers \citep[e.g.][]{eich57,jeff,eich63,deve}. The main motivation for
selecting this functional form was probably its mathematical simplicity, but also the fact that
some common kinds of distortions, such as rotation, scale and tilt are represented by monomials
in field coordinates. In the following, we will use a model currently considered for the JMAPS
astrometric satellite, but the techniques developed in this paper can be applied to any
variety of polynomial distortion models. The JMAPS model is:

\begin{eqnarray}
\label{poly.eq}
\xi&=& a_0  x+a_1 y+a_2+a_6 x^2+a_7 xy+a_8 r^2+a_{10}xr^2\\ \nonumber
\eta&=& a_3  x+a_4 y+a_5+a_6 xy+a_7 y^2+a_9 r^2+a_{10}yr^2
\end{eqnarray}
with $r$  denoting the radius, $r=\sqrt{x^2+y^2}$, for simplicity. This model includes the shifts
in $x$ and $y$ (model parameters $a_2$ and $a_5$), rotation, shear and scale ($a_0$ through $a_3$),
tilt  ($a_6$ and $a_7$), cubic radial distortion ($a_{10}$) and differential radial quadratic distortion (
$a_8$ and $a_9$).  It  is convenient to represent this 2D model in vector and matrix forms,
by stacking all the $x$ and $y$ measurements in a $2\times n$ column vector (where $n$ is the number
of reference stars), and likewise the standard coordinates $\xi$ and $\eta$, and arranging  the
model term values in a $2n$ by 11 matrix, so that the least-squares problem takes the form
\eb
\label{model.eq}
\left[
\begin{array}{ccccccccccc}
 {\bf x} & {\bf y}& {\bf 1}& {\bf x}^2&  {\bf xy}& {\bf r}^2& {\bf xr}^2& {\bf 0}& {\bf 0}& {\bf 0}& {\bf 0}\\
 {\bf 0}& {\bf 0}& {\bf 0}& {\bf xy} & {\bf y}^2& {\bf 0}& {\bf yr}^2&  {\bf x}& {\bf y}& {\bf 1}& {\bf r}^2
\end{array}\right] 
\:{\bf a}={\bf F}\: {\bf a}=\left[
\begin{array}{c} {\bf\xi}\\  {\bf\eta}  \end{array}
\right] 
\ee
where \eb  {\bf a}=\left[ \begin{array}{ccccccccccc}
a_0 & a_1 & a_2 & a_6 & a_7 & a_8 & a_{10} & a_3 & a_4 & a_5 & a_9 \end{array}\right]^{\rm T}.\nonumber\ee
The system of linear equations on a discrete set of data points $(\xi_i,\eta_i)$ is solved by the
least squares method yielding a solution
\eb
{\bf \tilde{a}}=({\bf F}^{\rm T}\;{\bf F})^{-1}\;{\bf F}^{\rm T}\;
\left[\begin{array}{c} {\bf\xi}\\  {\bf\eta}  \end{array}\right].
\ee
We are concerned with the propagation of the measurement error in the least squares solution.
Certainly, it depends on the distribution of data points in the focal plane, among other factors.
Reference stars in the focal plane can be non-uniformly distributed, for example, clustered
in the center of the field, or in one of the corners, giving rise to large errors in some of
the parameters, and to enhanced covariances between distortion coefficients. In space
astrometry projects, however, a single set of distortion coefficients is determined for
a large collection of frames, representing a large sample of random stellar configurations
and accruing a sufficiently high number density of reference stars. It is therefore  justified for practical
and theoretical reasons to consider an idealized sampling of the focal plane
with a large and dense regular grid of data points. The inner product of two vectors in that
case is replaced with its form for a Hilbert function space. For example, the inner product of
the fourth and the sixth columns of matrix ${\bf F}$ in Eq.~\ref{model.eq} is defined as
\eb
({\bf f}_4,{\bf f}_6)=\int_{-1}^{1}dx\int_{-1}^{1}  x^2(x^2+y^2)dy=\frac{56}{45}.
\ee
We assumed in this equation (without a loss of generality) that the detector area is square
and the field coordiantes are normalized to unity.

In practice, when an astrometric instrument is calibrated by its own observations, it is convenient
to simplify the transformations between the angular coordinates $(\xi,\eta)$ and the detector
coordinates $(x,y)$ to a simple form, e.g., a gnomonic projection plus a nominal scale conversion.
The corrections to the nominal parameters and the higher order terms, which may be time-dependent,
are sought for as small corrections to the detector coordinates, $(\Delta x,\Delta y)$. The
distortion coefficients are small enough to use the same model (such as in Eq. ~\ref{poly.eq})
in its differential form, by replacing $\xi$ and $\eta$ on the left-hand side with $\Delta x$
and $\Delta y$, respectively \citep{hilt, gre}. Once the small corrections to the nominal distortion
model are known, both forward and backward transformations between the field coordinates can be
easily performed. The JMAPS data reduction system is using this approach too, but we retain
the notations of Eq. ~\ref{poly.eq} in this paper for simplicity.

\section{Error propagation}

An error in the right-hand part of Eq.~\ref{model.eq} (i.e., measurement error $\epsilon$) propagates linearly
into the least-squares solution $\tilde{\bf a}$, so that
\eb
\label{error.eq}
\Delta\tilde{\bf a}={\bf F}^\dagger \epsilon
\ee
where  ${\bf F}^\dagger$ is the pseudoinverse of ${\bf F}$. Therefore, the error in the model
parameters is defined not only by the measurement error $\epsilon$, but also
by the properties of the design matrix, i.e., the composition of distortion terms.
Modeling field-dependent distortions becomes an important issue defining to some degree
the resulting astrometric accuracy. \citet{eich63} derived the  covariances of plate constants
for a series of commonly used models for the case of random uncorrelated errors ${\bf \epsilon}$.
They realized that the uncertainty of field characterization in the presence of random error
may be a complex, but quite predictable, function of field coordinates. The uncertainties tend
to become larger and more complex for more sophisticated polynomial models with higher
order terms. They also showed that the residual field-dependent error (which they called
systematic error) can be larger than  the gain in precision for unnecessarily involved
distortion models.

Correlations between some of the terms of a polynomial model are the main reason for this
potentially harmful feature. A large correlation between two distortion terms
should always be a concern, because the commonly used performance statistics based on
post-fit residuals, such as reduced $\chi^2$, do not capture the intrinsic statistical
dependencies within the solution. Astrometrists rarely know their instruments well
enough in advance, to the extent that a priori distortion  models can be used without some
experimentation or verification on real data. Even then, it is not clear where to stop
in building up a distortion model, because each additional term usually results in
smaller residuals, and the solution value can be deceptively large, indicating a
significant term, being in fact correlated with a number of other terms.

For these reasons, the use of orthogonal functions instead of algebraic polynomials
has long been advocated in the literature. \citet{bro89}, followed by \citet{bie93},
explained  the adverse effects of correlated terms, and considered the ease of
determination of a minimal-sufficient model to be the main advantage of orthogonalized
designs. Having determined the level of statistically significant reduction in
$\chi^2$, one can, for example, remove orthogonal terms from the model in turns,
until this level is achieved with the smallest number of terms.

The effects of systematic errors have largely been unheeded in the literature. By a systematic
error we understand any deterministic perturbation of the data which can be expressed
as a function of field coordinates. Obviously, there is an infinite set of possible
systematic errors, and analysis of their propagation appears to be intractable. Beyond
the common systematic errors, such as a focal length variation, these effects are
hard to predict in complex astrometric instruments. However, instead of guessing possible
systematic errors and simulating their effects, one can pose this problem: is there a
specific systematic error which propagates into the least squares solution with the
largest gain? An astonishingly simple answer is obtained through the Singular Value Decomposition
(SVD) technique \citep{gol}. Let
\eb
{\bf F}={\bf  U\;\Sigma\;V}^{\rm T}
\ee
be the SVD of the design matrix ${\bf F}$. The LS solution is
\eb
{\bf \tilde{a}}={\bf V}\;{\bf \Sigma}^{-1}\;{\bf U}^{\rm T}\;
\left[\begin{array}{c} {\bf\xi}\\  {\bf\eta}  \end{array}\right].
\ee
For an overdetermined system with $k$ model terms (i.e., of rank k), ${\bf \Sigma}^{-1}$
is a diagonal matrix with $k$ nonzero (positive) numbers in the diagonal in increasing order.
Therefore,  in the projection of the data vector onto the basis, ${\bf U}^{\rm T}\,
[{\bf\xi}\:  {\bf\eta}]^{\rm T}$, only the first $k$ columns of the orthogonal matrix 
${\bf U}$ contribute to the solution. In other words, the first $k$ columns of
${\bf U}$ define the basis of the subspace spanned by ${\bf F}$ and the remainder is
the basis of its null space\footnote{In Matlab, the significant part of SVD is realized
through the economic SVD: {\tt [u,sigma,v]=svd(f,0)}.}. The LS solution is impervious
to any perturbation of data in the null space. If $\epsilon$ is a systematic perturbation
of the data, the squared norm of the resulting systematic error of the solution is
\eb
\|\delta \|^2=\delta^{\rm T}\delta=\epsilon_U^{\rm T}\;{\bf \Sigma}^{-2}\;\epsilon_U
\ee
where $\epsilon_U={\bf  U}^{\rm T}\epsilon$ is the projection of $\epsilon$ onto the basis of
${\bf F}$. Recalling that ${\bf \Sigma}={\rm diag}([\sigma_1\; \sigma_2\;\ldots\;\sigma_k]^{\rm T})$
and $\sigma_1>\sigma_2>\ldots >\sigma_k$, the maximum $\|\delta \|^2$  is achieved when
$\epsilon$ is aligned with ${\bf u}_k=U(:,k)$. This follows from the constrained Lagrange
problem, where the Lagrange multiplier is one of the $\sigma_i^{-2}$ and among all $\epsilon$
of unit norm the largest solution error happens when $\epsilon_U={\bf u}_k^{\rm T}\;\epsilon$,
hence, $\epsilon={\bf u}_k$.

\begin{figure}[htbp]
  \centering
  \includegraphics[angle=0,width=0.7\textwidth]{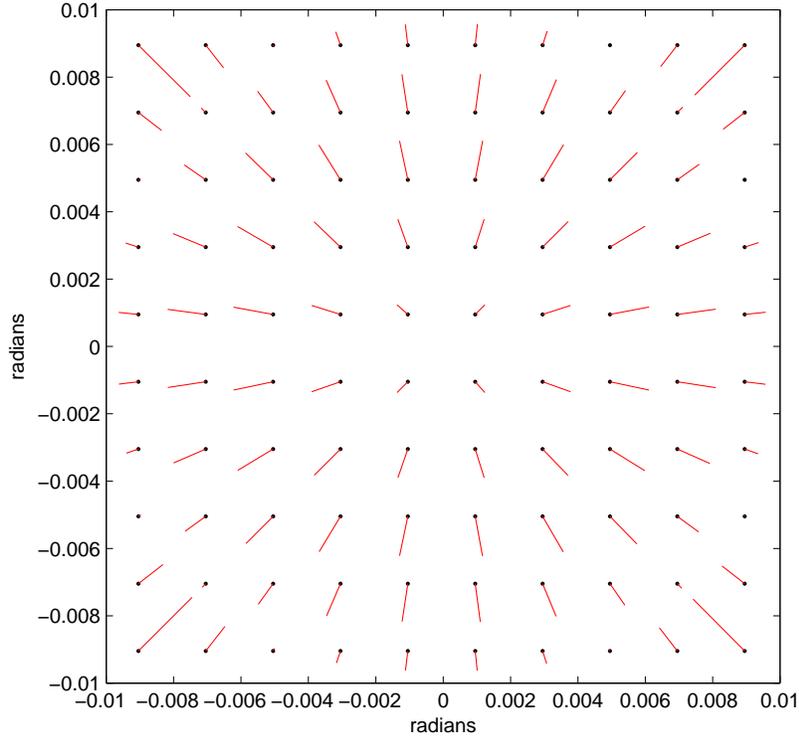}
\caption{The worst field distortion for the nominal JMAPS distortion, which yields
the largest error of the focal plane parameters. \label{worst.fig}}
\end{figure}

Thus, the $k$th long singular vector, which is the \textsl{least significant principal component}
of model ${\bf F}$, is the worst systematic error among all possible systematic perturbations of unit norm.

\section{The case of JMAPS}
Using a dense grid of points on the unit square, the SVD of the nominal JMAPS distortion model
(Eq. \ref{model.eq}) can be computed and the worst systematic error numerically derived.
The result is a higher order field distortion shown in Fig.~\ref{worst.fig}. It is similar
to the scale in the center of the field of view, but instead of growing linearly with radius,
it turns to zero and reverses the sign in the outer regions. Recalling that the worst
perturbation is one of the orts of the subspace spanned by the columns of ${\bf F}$,
it can be expressed as a certain linear combination of the model functions. In this
case, apparently, the differential scale terms $[{\bf x}\;{\bf 0}]^{\rm T}$ and
$[{\bf 0}\;{\bf y}]^{\rm T}$ in combination with the radial distortion term $[{\bf xr}^2\;{\bf yr}^2]^{\rm T}$
are the dominating contributors to the worst perturbation ${\bf u}_{11}$.

How does one estimate the impact of the worst systematic error? Clearly, the term ``worse" or ``bad" can only be
meaningful in a relative, or comparative sense. The singular values readily provide
a meaningful comparison of various possible systematic errors. The most benign systematic
perturbation is ${\bf u}_1$, and the ratio of the smallest possible to the largest possible
systematic error $\|\delta\|$ is simply $\sigma_k/\sigma_1$, which is the condition number
of ${\bf F}$. A very small condition number indicates an ill-conditioned problem.
An inadeptly chosen distortion model may have a deficient rank, in which case one or
more singular values equal zero, and the solution error can be infinite. For the
JMAPS model in Eq.~\ref{model.eq}, the condition number is $0.14652$, indicating
a moderately conditioned model. The worst unit-norm perturbation results in a
systematic error whose norm is 7 times that of the most benign unit-norm perturbation
from the subspace of ${\bf F}$. 

\section{Orthogonal distortion models on the unit disk} 
\begin{deluxetable}{lr|lr}
\tablecaption{Algebraic polynomials and Zernike function \label{zer.tab}}
\tablewidth{0pt}
\tablehead{
\multicolumn{2}{c}{Zernike to Polynomials conversions} & \multicolumn{2}{c}{Polynomials to
Zernike conversions} \\ \hline
}
\startdata
$Z_0^{0} $ & $1$ & $1$ \dotfill& \dotfill$Z_0^{0}$ \\ [1ex]
$Z_1^{-1} $ & $2r\sin\phi$ & $x$ \dotfill& \dotfill$\frac{1}{2}Z_1^{-1}$ \\ [1ex]
$Z_1^{+1} $ & $2r\cos\phi$ & $y$ \dotfill& \dotfill$\frac{1}{2}Z_1^{+1}$ \\ [1ex]
$Z_2^{0} $ & $\sqrt{3}(2r^2-1)$ & $x^2+y^2$ \dotfill& \dotfill$\frac{1}{2\sqrt{3}}Z_2^{0}+\frac{1}{2}Z_2^0$ \\ [1ex]
$Z_2^{-2} $ & $\sqrt{6}r^2\sin2\phi$ & $xy$ \dotfill& \dotfill$\frac{1}{2\sqrt{6}}Z_2^{-2}$ \\ [1ex]
$Z_2^{+2} $ & $\sqrt{6}r^2\cos2\phi$ & $y^2$ \dotfill& \dotfill$\frac{1}{2\sqrt{6}}Z_2^{+2}+\frac{1}{4\sqrt{3}}Z_2^{0}
+\frac{1}{4}Z_0^{0}$ \\ [1ex]
$Z_3^{-1} $ & $2\sqrt{2}(3r^3-2r)\sin\phi$ & $x(x^2+y^2)$ \dotfill& \dotfill$\frac{1}{6\sqrt{2}}Z_3^{-1}+\frac{1}{3}Z_1^{+1}$ \\ [1ex]
$Z_3^{+1} $ & $2\sqrt{2}(3r^3-2r)\cos\phi$ & $y(x^2+y^2)$ \dotfill& \dotfill$\frac{1}{6\sqrt{2}}Z_3^{+1}+\frac{1}{3}Z_1^{+1}$ \\
\enddata
\end{deluxetable}

The two-dimensional Zernike functions (also known as the circle polynomials of Zernike) provide
a ready-to-use orthogonal basis of scalar functions on the unit disk. These functions are
used to decompose the optical aberration function \citep{bor}, and find other numerous applications
in optics and engineering. Table \ref{zer.tab} lists several low-order terms starting with
$Z_0^{0} $, which is a constant. An expanded set of the Zernike functions and transformations of
algebraic monomials is given by \citet{math}. The Zernike functions are normalized to $\sqrt{\pi}$
with a weight $r$:
\eb
\int_0^1 \int_0^{2\pi} Z_m^k Z_n^l r\;dr\,d\phi=\pi \delta_{mn} \delta_{kl}, \nonumber
\ee
and each of the algebraic monomials in Eq. \ref{model.eq} is uniquely represented by a linear combination
of Zernike functions. The polynomial model is not orthogonal because some of the terms include the
same Zernike function, as can be seen in Table \ref{zer.tab}. For example, the product of the 6th term
(with $a_6$) and 0th term (with $a_0$) is equal to $\pi/6$. One can replace the non-orthogonal terms
in model (\ref{model.eq}) with their significant Zernike counterparts, arriving at
\eb
\label{mod2.eq}
\left[
\begin{array}{ccccccccccc}
 {\bf Z}_1^{-1} & {\bf Z}_1^{+1}& {\bf Z}_0^{0}& \frac{1}{\sqrt{2}}{\bf Z}_2^{+2}&  
 \frac{1}{\sqrt{2}}{\bf Z}_2^{-2}& {\bf Z}_2^{0}& \frac{1}{\sqrt{2}}{\bf Z}_3^{-1}& {\bf 0}& {\bf 0}& {\bf 0}& {\bf 0}\\
 {\bf 0}& {\bf 0}& {\bf 0}& \frac{1}{\sqrt{2}}{\bf Z}_2^{-2} & \frac{1}{\sqrt{2}}{\bf Z}_2^{+2}& 
 {\bf 0}& \frac{1}{\sqrt{2}}{\bf Z}_3^{+1}&  {\bf Z}_1^{-1}& {\bf Z}_1^{+1}& {\bf Z}_0^{0}& {\bf Z}_2^{0}
\end{array}\right] 
\:{\bf a}=\left[
\begin{array}{c} {\bf\xi}\\  {\bf\eta}  \end{array}
\right] 
\ee
This is the closest to the original (\ref{model.eq}) model, which is orthogonal on the unit disk.
It can be easily expanded to higher orders by including more terms constructed from the Zernike functions.

\section{Orthogonal distortion models on the unit square}
\label{square.sec}
Instruments of optical astrometry are equipped today with CCD detectors or other electronic sensors 
that are rectangular in shape. It is practical to use a functional basis, which is orthonormal on
a quadrangle. Without a loss of generalization, a basis on the unit square can be applied to any
rectangular field of view by means of coordinate normalization. The Zernike functions described
in the previous section are not relevant, because some of them are not mutually
orthogonal on the unit square.

We demonstrate herewith, how a functional basis can be constructed on the unit square starting with
a model, which includes polynomials or Zernike terms. This is implemented by the Gram-Schmidt orthogonalization
process in the space of 2D scalar functions. If $\bf{Y}_m$ are the original (non-orthogonal) vector
functions, and
$\bf{V}_m$ are the new basis functions orthogonal on the square, the algorithm is:
\begin{eqnarray}
\label{gramsch.eq}
\bf{V}_1 &=& \bf{Y}_1 \nonumber\\
\bf{V}_2 &=& \bf{Y}_2-(\bf{V}_1,\bf{Y}_2)/(\bf{V}_1,\bf{V}_1)\,\bf{V}_1 \\
\bf{V}_3 &=& \bf{Y}_3-(\bf{V}_1,\bf{Y}_3)/(\bf{V}_1,\bf{V}_1)\,\bf{V}_1-(\bf{V}_2,\bf{Y}_3)
/(\bf{V}_2,\bf{V}_2)\,\bf{V}_2 \nonumber\\
&&\ldots \nonumber
\end{eqnarray}
The inner products in this case are non-weighted integrals over the unit quadrangle, i.e.,
\eb
(\bf{V}_m,\bf{V}_n)=\int_{-1}^{+1} \int_{-1}^{+1} \bf{V}_m\cdot \bf{V}_n \;dx\,dy. \nonumber
\ee
Normalized functions $\hat{V}_m=V_m/\sqrt{(V_m,V_m)}$ have the additional advantage of all singular
values being equal to unity for a dense and uniform set of sample points, which makes a perfectly conditioned
model.

A set of orthogonal functions $\bf{V}_m$, $m=1,\ldots,12$, derived from the original JMAPS model (\ref{poly.eq}) is given in Table~\ref{vfunk.tab}, along with their norms on the unit square. The updated model of JMAPS distortions in
terms of orthogonal functions is
\eb
\label{modV.eq}
\left[
\begin{array}{cccc}
 \hat{\bf V}_1 & \hat{\bf V}_2 & \ldots & \hat{\bf V}_l
\end{array}\right] 
\:{\bf a} = \left[
\begin{array}{c} {\bf\xi}\\  {\bf\eta}  \end{array}
\right]
\ee
As long as the field of view is well sampled with observations of calibration stars, the design matrix
in these equations is nearly orthogonal, i.e., ${\bf F}^T {\bf F}\simeq {\bf I}_k$.

\section{Discussion}
Using the normalized vector fields based on the 2D Zernike polynomials for circular fields of view,
Eq. \ref{mod2.eq}, or 2D orthonormal polynomials for rectangular fields of view, Eq. \ref{modV.eq},
leads to the most stable and well-conditioned solutions for instrument calibration parameters.
Any systematic error of unit norm will result in a solution error of the same magnitude. A random
error of unit weight ($\epsilon \; \epsilon^{\rm T}={\bf I}$, per Eq. \ref{error.eq}) will result
in an uncorrelated set of model coefficients, each having the same error expectancy, because
the covariance of the LS solution is close to unity. This property of orthonormal
distortion models comes in handy when we have to assess the significance of different terms
in our model. On the one hand, the insignificant terms, which describe certain types of
field dependent corrections that are not really present in the data, should be identified
and eliminated as soon as possible, because the redundant degrees of freedom degrade the overall
astrometric solution. This is especially important when the instrument is not very stable,
resulting in numerous field-dependent calibration unknowns in the global adjustment.
On the other hand, if the observational residuals prove too large, and their distribution
suggests the presence of higher order terms, which are not captured by the distortion model,
the orthonormal basis should be built up using the generalized Gram-Schmidt algorithm (\S 
\ref{square.sec}). 

Orthogonal plate models are also useful when possible
internal degeneracies in the instrument characterization have to be identified. 
A specific, but very common case
is when the focal plane assembly includes more than one photodetector array, each being
characterized by its own set of calibration parameters. In such hierarchical calibration
models, lower-level parameters may be strongly correlated or completely degenerate with the
higher-level parameters. For example, a shift of each CCD in the same direction by the same
amount is indistinguishable from a common shift of the entire focal plane. Such double
accounting of the same effect should be avoided, because it can lead to a complete
solution failure, or divergence of iterated solutions. With an orthonormal plate model,
the degeneracies are simply quantified by projection of the additional terms onto the
previously accumulated basis. If the projections turn out to be significantly nonzero,
the new terms can be orthogonalized analytically, as described in this paper, or
numerically by the principal component method.
 
\begin{deluxetable}{l |r}
\tablecaption{Vector basis functions orthogonal on the unit square \label{vfunk.tab}}
\tablewidth{0pt}
\tablehead{
\colhead{Function} & \colhead{Square norm}
}
\startdata
$\bf{V}_1=\left[\begin{array}{c} \bf{1}\\ \bf{0} \end{array}\right]$ 
		&	$(\bf{V}_1,\bf{V}_1)=4$ \\ [1.8ex]
$\bf{V}_2=\left[\begin{array}{c} \bf{0}\\ \bf{1} \end{array}\right]$ 
		&	$(\bf{V}_2,\bf{V}_2)=4$ \\ [1.8ex]
$\bf{V}_3=\left[\begin{array}{c} \bf{x}\\ \bf{0} \end{array}\right]$ 
		&	$(\bf{V}_3,\bf{V}_3)=\frac{4}{3}$ \\ [1.8ex]
$\bf{V}_4=\left[\begin{array}{c} \bf{0}\\ \bf{y} \end{array}\right]$ 
		&	$(\bf{V}_4,\bf{V}_4)=\frac{4}{3}$ \\ [1.8ex]
$\bf{V}_5=\left[\begin{array}{c} \bf{y}\\ \bf{0} \end{array}\right]$ 
		&	$(\bf{V}_5,\bf{V}_5)=\frac{4}{3}$ \\ [1.8ex]
$\bf{V}_6=\left[\begin{array}{c} \bf{0}\\ \bf{x} \end{array}\right]$ 
		&	$(\bf{V}_6,\bf{V}_6)=\frac{4}{3}$ \\ [1.8ex]
$\bf{V}_7=\left[\begin{array}{c} \bf{x}^2-\frac{1}{3}\\ \bf{xy} \end{array}\right]$ 
		&	$(\bf{V}_7,\bf{V}_7)=\frac{4}{5}$ \\ [1.8ex]
$\bf{V}_8=\left[\begin{array}{c} \bf{xy}\\ \bf{y}^2-\frac{1}{3} \end{array}\right]$ 
		&	$(\bf{V}_8,\bf{V}_8)=\frac{4}{5}$ \\ [1.8ex]
$\bf{V}_9=\left[\begin{array}{c} \bf{xr}^2-\frac{14}{15}\bf{x}\\ \bf{yr}^2-\frac{14}{15}\bf{y} \end{array}\right]$ 
		&	$(\bf{V}_9,\bf{V}_9)=\frac{1984}{4725}$ \\ [1.8ex]
$\bf{V}_{10}=\left[\begin{array}{c} \frac{5}{9}\bf{x}^2+\bf{y}^2-\frac{14}{27}\\ -\frac{4}{9}\bf{xy} \end{array}\right]$ 
		&	$(\bf{V}_{10},\bf{V}_{10})=\frac{77344}{164025}$ \\ [1.8ex]
$\bf{V}_{11}=\left[\begin{array}{c} -\frac{4}{9}\bf{xy}\\ \bf{x}^2+\frac{5}{9}\bf{y}^2+-\frac{14}{27} \end{array}\right]$ 
		&	$(\bf{V}_{11},\bf{V}_{11})=\frac{77344}{164025}$ \\ [1.8ex] \hline
$\bf{V}_{12}=\left[\begin{array}{c} \frac{35}{62}\bf{x}^3-\frac{27}{62}\bf{xy}^2-\frac{30}{155}\bf{x}\\ 
\frac{35}{62}\bf{y}^3-\frac{27}{62}\bf{x}^2\bf{y}-\frac{30}{155}\bf{y} \end{array}\right]$ 
		&	$(\bf{V}_{12},\bf{V}_{12})=\frac{16}{155}$
\enddata
\end{deluxetable}

\end{document}